\documentclass[preprint,floatfix,aps,prb,showpacs]{revtex4}  
\usepackage{graphicx}

\def\o0{\omega_0}

\def\ome{\omega}
\def\ome0{\omega_0}
\def\vep{\varepsilon}
\def\wh{\widehat}
\def\P0{\wh{\cal P}_0}

\def\eps{\varepsilon}
\newcommand{\beq}{\begin{equation}}
\newcommand{\eeq}{\end{equation}}
\newcommand{\beqa}{\begin{eqnarray}}
\newcommand{\eeqa}{\end{eqnarray}}

\begin{document}
\title{Theory of electronic transport through a triple quantum dot in the presence of magnetic field}  
  
\author{F. Delgado}  
\affiliation{Quantum Theory Group, Institute for Microstructural Sciences,  
       National Research Council, Ottawa, Ontario, Canada K1A 0R6}  
\affiliation{Department of Physics, University of Ottawa,  MacDonald Hall, 150 Louis Pasteur,
Ottawa, Ontario, Canada K1N 6N5}    
  
\author{P. Hawrylak}  
\affiliation{Quantum Theory Group, Institute for Microstructural Sciences,  
        National Research Council, Ottawa, Ontario, Canada K1A 0R6}

\begin{abstract}  
Theory of electronic transport through a triangular triple quantum dot subject to a 
perpendicular magnetic field is developed using a tight binding model. We show that
magnetic field allows to engineer degeneracies in the triple quantum dot energy spectrum. 
The degeneracies lead to zero electronic transmission and sharp dips in the current
whenever a pair of degenerate states lies between the chemical potential of the two leads.
These dips can occur with a periodicity of one flux quantum if only two 
levels contribute to the current or with half flux quantum if
the three levels of the triple dot contribute.
 The effect of strong bias 
 voltage and different lead-to-dot connections on Aharonov-Bohm
  oscillations in the conductance
  is also discussed.  
\end{abstract}  
  
  
\maketitle

\section{Introduction}  
Using charge sensing techniques  Gaudreau {\em et al.}\cite{gaudreau_studenikin_prl2006,korkusinski_gimenez_prb2007} recently 
demonstrated lateral triple quantum dot (TQD) molecule with 
controlled number of electrons, down to zero.  
Preliminary transport experiments  in external magnetic
field\cite{gaudreau_sachrajda_icps2006,ihn_sigrist_njphys2007} 
showed signatures 
of Aharonov-Bohm (AB) oscillations, indicating 
coherent coupling between the constituent dots.
Motivated by forthcoming experiments,  we present here  a theory describing signatures
of AB oscillations in 
transport through the TQD in a perpendicular
magnetic field. 
   Using tight binding model we show that  
magnetic field allows us to engineer  degeneracies in the triple quantum dot spectrum, and
that these degeneracies lead to zero electronic transmission and to sharp dips in the current.
These anomalies in transport can appear with different periodicities or be suppressed depending
on the applied source-drain voltage and dot energies.
The main features of the transport
are explained as an interplay between Fano resonances and AB oscillations. 
The AB oscillations apparent in the conductance allows
for unambiguous identification of TQD parameters.
The effects of strong bias voltage on the conductance are also discussed. 
Two different lead-to-dot connections are considered: a left lead connected to a single
dot and left lead connected to two dots. The first configuration leads
to a periodic oscillation of the current with the magnetic field while the second one
breaks the periodicity introducing an extra structure superimposed on the
oscillatory behaviour as a function of the magnetic flux.

 In our tight-binding model, effects associated
 with the electron-electron interactions, extensively analyzed in the context of
 transport through single quantum dots,
 experimentally\cite{gordon_gores_prl_1998,cronenwett_oosterkamp_science_1998,schmid_weis_phys_1998}
 and theoretically,\cite{hewson_book_1993,hershfield_davies_prl_1991,
meir_wingreen_prl_1992,jauho_wingreen_prb_1994,yeyati_rodero_prl_1993,hettler_schoeller_prl_1995}  
in relation with the Kondo physics do not appear.
However, broadening of molecular energy levels is properly
taken into account in our model.
A perpendicular magnetic field is accounted for by Peierls phase 
factors\cite{peierls_zphys1933,luttinger_pr1951} in the
single-particle tunneling elements,
leading to AB oscillations in the conductance with period of one flux quantum 
$\Phi_0=e/hc$ ($e$- electron charge, $h$-Plank's constant and $c$- speed of light),
and anomalies at half flux quantum.
The AB oscillations in the conductance are inherent to rings threaded by magnetic 
flux.\cite{gefen_imry_prl1983,buttiker_imry_pra1984}
Flux period of $\Phi_0$ is observed
in conventional AB experiments with electrons propagating in field-free regions\cite{chambers_prl_1960,tonomura_matsuda_prl_1982}
 and also in mesoscopic experiments,
for example in metal rings\cite{web_washburn_prl_1985}
 or in electronic Mach-Zehnder 
 interferometers.\cite{yacovy_heiblum_prl_1994,ji_chung_nature_2003}
Furthermore, $\Phi_0/2$ periods can be also observed due to
weak localization effects.\cite{altshuler_aronov_jetp_1981,chandrasekhar_rooks_prl_1985}

At difference with previous works on equilateral triple dot connected to 
leads where only the linear response to a small bias was analyzed,\cite{ingersent_ludwig_prl2005,kuzmenko_kikoin_prl2006,jiang_sun_jphysc2007} or works
based on a master equation approach to a single electron 
tunneling\cite{groth_michaelis_prb2006,michaelis_emary_euphyl2006,emary_condmatt2007} valid
only in the limit of large applied bias,
we discuss the differential conductance in the case of arbitrary applied bias voltage
and magnetic field in an exact non-perturbative way, including the experimental
 conditions in Ref. ~\onlinecite{gaudreau_sachrajda_icps2006}.

%%%%%%%%%%%%% organization of paper %%%%%%%%%%%%%%%%%%%%%%%%%%

The paper is organized as follows.
In Sec. \ref{model} we introduce the Hamiltonian describing the system
while Sec. \ref{transfer} explains how to obtain the transmission coefficient
from the transfer matrix and the scattering boundary conditions.
The AB oscillations in the current are analyzed in Sec. \ref{Fanores} together
with the Fano line-shape of the transmission probability
 while the anomalous behaviour 
of the transmission close to multiples of half flux quantum is studied in 
Sec. \ref{degenerate}.
 The conductance in the non-linear regime is analyzed 
in Sec. \ref{results}.  The paper is summarized
in Sec. ~\ref{conclusions}.

\section{Model\label{model}}
\begin{figure}
\begin{center}
\includegraphics[angle=0,width=0.7\linewidth]{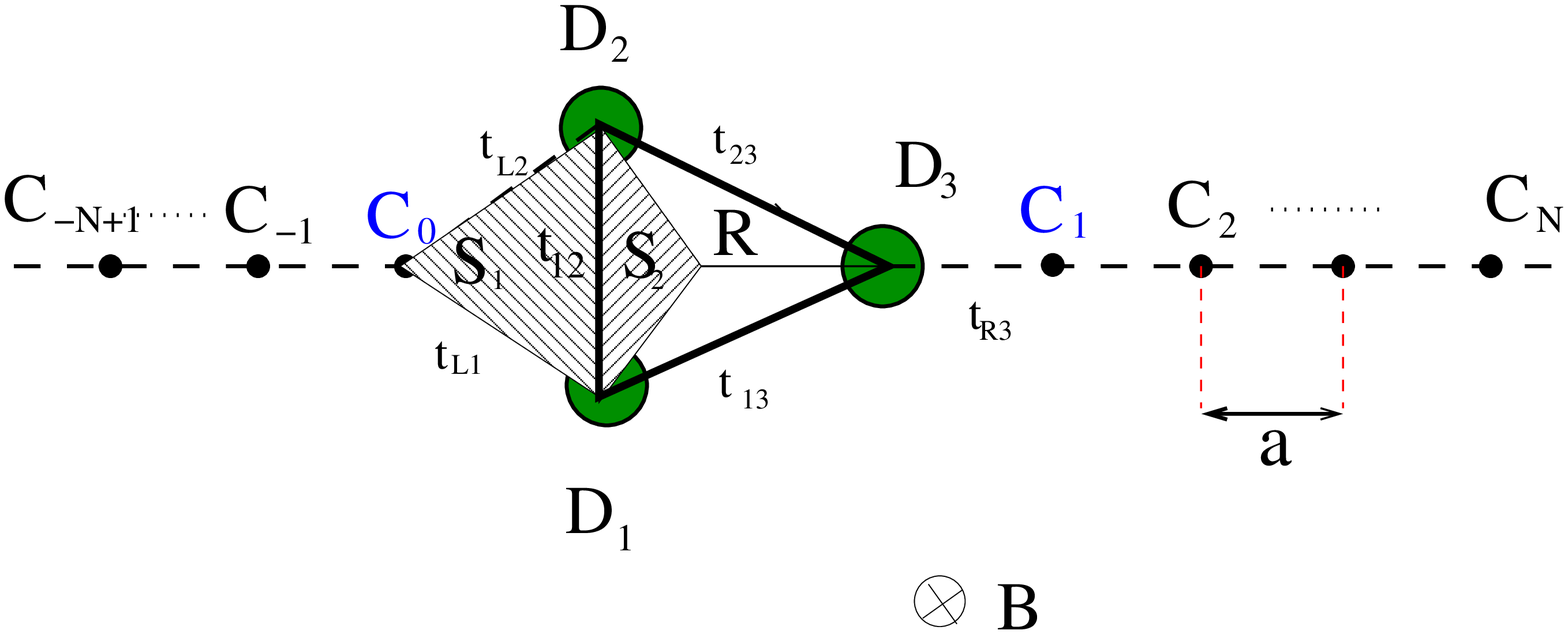}
\end{center}
\caption{Schematic diagram of the spatial layout of
the triple dot and the two leads. Allowed tunneling between
different sites is marked with thick long-dashed lines.}
\label{tqdf_1}
\end{figure}

The triple dot connected to leads is plotted schematically in
Fig. \ref{tqdf_1}. The leads are described within a one-dimensional 
tight-binding model, with nearest neighbors hopping $t_L$.
Each dot is represented by a single orbital,
connected to nearest neighbors by magnetic field dependent
hopping matrix elements $t_{ij}(B)$, with 
$i\;,j\;=1,2,3$ ( $i\ne j$). The left lead is connected to the dots 1 and 2, 
see Fig. \ref{tqdf_1}, through the hopping elements $t_{L1}$ and $t_{L2}$,
while the right lead is connected only to dot 3 with hopping matrix element
$t_{R3}$. The TQD is subject to a uniform perpendicular magnetic field ${\bf B}$,
${\bf B}=B {\bf \hat{z}}$.
The Hamiltonian describing the system is then given by
\beqa
H=H_{TQD}+H_{leads}+H_{LD},
\label{htot}
\eeqa
where $H_{TQD}$ is the Hamiltonian corresponding to an electron
in an isolated triple dot
\beqa  
  H_{TQD}=   
  \sum\limits_{i=1}^{3}  \left(E-\Delta V/2\right) d_{i}^+ d_{i}  
+ \sum\limits_{i,j=1,i\neq j}^{3}   
             t_{ij}(B) d_{i}^+ d_{j}, 
 \nonumber
\\
\label{htqd}  
\eeqa  
the operators $d_{i}$ ($d_{i}^+$) annihilate 
(create) an electron  in dot $i$. $E$ is the energy level of each quantum dot and
$\Delta V$ is the energy bias between the two leads. 
Notice that as a first order approximation, we have assumed that the shift
in the dot energy levels as a function of the applied bias is the same 
for all dots, $-\Delta V/2$.
Furthermore, for identical dots the hopping matrix elements at $B=0$ satisfy
$t_{ij}=t\;\: \forall \;i,j$.

\begin{figure}
\begin{center}
\includegraphics[angle=0,width=0.4\linewidth]{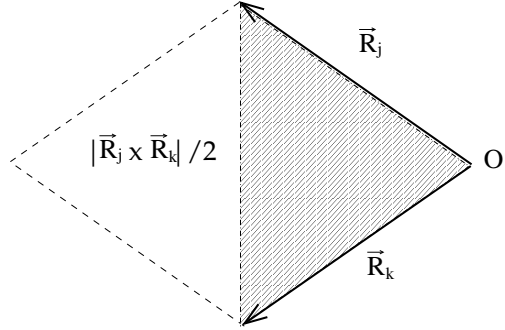}
\end{center}
\caption{Area responsible of the phase difference between two points
${\bf R}_j$ and ${\bf R}_k$ when a vector potential ${\bf A}$ with
a gauge centered in the point $O$ is considered.}
\label{fig2_0}
\end{figure}

$H_{leads}$ is the Hamiltonian 
describing the two non-interacting leads with $N$ sites each,
\beqa
H_{leads}&=&\epsilon_L c_0^+c_0+\sum_{i=-N+1}^{-1}\left[ \epsilon_{L} c_{i}^+ 
c_{i}+
t_L \left(c_{i}^{+}c_{i+1}+c_{i+1}^+c_{i}\right)\right]
\cr\cr
&+& \epsilon_R c_1^+c_1+\sum_{i=2}^{N}\left[ \epsilon_R c_{i}^+ 
c_{i}+
t_L \left(c_{i-1}^{+}c_{i}+c_i^+c_{i-1}\right)\right],
\label{hlead}
\eeqa
where $c_i^+$ and $c_i$ are respectively the creation and 
annihilation operators of an electron
on site $i$ in the leads, $\epsilon_{L}$ is the on-site energy in the
leads at zero bias and $\epsilon_R=\epsilon_L-\Delta V$.
Both leads are characterized by the
same hopping matrix elements, $t_L$.
Finally, the interaction Hamiltonian
$H_{LD}$ is given by 
\beqa
H_{LD}=t_{L1}(B)c_0^+d_1+t_{L2}(B)c_0^+d_2+t_{R3}(B)c_1^+d_3+hc.
%\sum_{i=1}^{2}\left[t_{L}^{(i)}(B) c_{0}^{+}d_{i}+hc.\right]+
%t_{R}^{(3)} \left(c_{1}^{+}d_{3}+hc.\right)
\label{hint}
\eeqa
The magnetic field $B$ renormalizes the single-particle tunneling elements $t_{jk}$ 
by Peierls phase factors,\cite{peierls_zphys1933,luttinger_pr1951}  
$t_{jk}(B)=t_{jk}e^{2\pi i \phi_{jk}}$, where 
$\phi_{jk}=\frac{e}{2\pi\hbar c}\int_{{\bf R}_j}^{{\bf R}_k} {\bf A}. d{\bf l}$. 
${\bf A}$ is the corresponding vector potential and ${\bf R}_j$ and ${\bf R}_k$ are
the positions of the sites connected by the hopping elements $t_{jk}$ 
Taking the 
symmetric gauge in which ${\bf A}=[-By,Bx,0]$,
 the phase difference between
two points ${\bf R}_j$ and ${\bf R}_k$ is given
by $\phi_{jk}=\frac{1}{2}{\bf B}\cdot\left({\bf R}_k\times {\bf R}_j\right)\Phi_0$, 
see Fig. \ref{fig2_0}, with 
$\Phi_0=\frac{hc}{e}$ the magnetic flux quantum.

For the three quantum dots located in the corners of an  
equilateral triangle we have $\phi_{12}=\phi_{23}=\phi_{31}=-\phi/3$. 
Here, $\phi=3\sqrt{3} B R^2 /4\Phi_0 $ is the number of magnetic 
flux quanta threading the area of the TQD, with
$R$ -the distance from the center of the triangle to each dot, 
identified in Fig.~\ref{tqdf_1}.   

For the general case where the left lead is connected to
dots one and two 
through the hopping matrix elements $t_{L1}(B)$ and $t_{L2}(B)$ respectively, 
while dot three is connected only to the right lead with hopping parameter
$t_{R3}$, as shown in Fig. \ref{tqdf_1}, there is an extra magnetic flux $\phi'$. In this case, 
$t_{L1}(B)=t_{L1}e^{2\pi i \phi'}$ and 
$t_{L2}(B)=t_{L2}e^{-2\pi i \phi'}$.
 If $S_1$ and $S_2$ are the shaded areas on
Fig. \ref{tqdf_1}, the two fluxes are related through the ratio of areas
$\phi'=-\phi\frac{4\pi^2}{3\sqrt{3}}\left(\frac{S_1+S_2}{\pi R^2}\right)$.
As it will be shown, these phases have an important effect on transport leading 
to a non-periodic behaviour of the transmission with the magnetic field, except 
for the particular case where
$\phi/\phi'$ is a rational number.

\section{Transfer matrix and scattering matrix\label{transfer}}
Our aim in the present section is to obtain a $2\times 2$ transfer 
matrix ${\cal T}$ which relates the 
amplitude of the wave functions on the last two sites of the left lead, $C_{-1}$ and 
$C_0$, with those at the first two positions of the right lead, $C_1$ and $C_2$. 
In matrix form,
\beqa
\left(\begin{array}{cc}
{\cal T}_{11} &{\cal T}_{12}\\
{\cal T}_{21} & {\cal T}_{22}
\end{array}\right).\left(\begin{array}{cc}
C_{-1}\\
C_0\end{array}\right)=
\left(\begin{array}{cc}
C_1\\
C_2
\end{array}\right).
\label{ttransfer}
\eeqa
Here and in the following sections, we will use the notation
$X_i \equiv \langle {\bf r}|x_i^+|0\rangle$ with $x_i^+=c_i^+,\;d_i^+$
for the amplitude of
the wave function at position $i$. 
The transmission and reflection coefficients can be then obtained by
imposing the scattering boundary conditions.
 In particular, if we consider a left 
incident plane wave with wavevector $k$, the amplitudes at the left and right of
the triple dot will be given by
\beqa
C_0 &=&1+R\cr
C_{-1}&=&e^{-ika}+Re^{ika}\cr
C_1&=&T\cr
C_2&=&Te^{ik'a},
\label{boundary}
\eeqa
The wavevector and the energy of the incident electron $\vep$ is related through the
dispersion relation in an infinite lead, 
$ka=\mbox{arc cos}\left(\frac{\vep-\epsilon_{L}}{2t_L}\right)$ and
$k'a=\mbox{arc cos}\left(\frac{\vep-\epsilon_{L}-\Delta V}{2t_L}\right)$, with 
$a$ the lattice constant.
 Then, from
Eqs. (\ref{ttransfer}) and (\ref{boundary}), reflection $R$ and transmission $T$
can be expressed as:
\beqa
R&=&\frac{e^{-ika}\left[-e^{ik'a}{\cal T}_{11}+{\cal T}_{21}+e^{ika}
\left(-e^{ik'a}{\cal T}_{12}+{\cal T}_{22}\right)\right]}
{e^{ik'a}\left(e^{ika}{\cal T}_{11}+{\cal T}_{12}\right)-\left(e^{ika}
{\cal T}_{21}+{\cal T}_{22}\right)},
\cr\cr
T&=&\frac{e^{-ika}\left(-1+e^{2ika}\right)
\left(-{\cal T}_{12}{\cal T}_{21}+{\cal T}_{11}{\cal T}_{22}\right)}
{e^{ik'a}\left(e^{ika}{\cal T}_{11}+{\cal T}_{12}\right)
-\left(e^{ika}{\cal T}_{21}+{\cal T}_{22}\right)}.
\label{trleft}
\eeqa
The transfer matrix ${\cal T}$ will be obtained by applying
 the Hamiltonian to the amplitudes.
It is convenient to express the original Hamiltonian (\ref{htot}) in the basis
of eigenfunctions of the isolated triple dot. If we define the 
annihilation operators $\bar{d}_1,\bar{d}_2, \bar{d}_3$ 
in terms of the corresponding annihilation operators for electrons
on sites 1, 2 and 3 as 
\beqa
\left\{
\begin{array}{lll}
d_1=\frac{1}{\sqrt{3}}\left(\bar{d}_1+\bar{d}_2+\bar{d}_3\right)\\
d_2=\frac{1}{\sqrt{3}}\left(\bar{d}_1+e^{-2\pi i/3}\bar{d}_2+e^{2 \pi i/3}\bar{d}_3\right)\\
d_3=\frac{1}{\sqrt{3}}\left(\bar{d}_1+e^{2\pi i/3}\bar{d}_2+e^{-2 \pi i/3}\bar{d}_3\right)\\
\end{array}
\right.,
\eeqa
the triple dot Hamiltonian will be diagonal at all values of the magnetic field:
\beqa
\bar{H}_{TQD}=\epsilon_1\bar{d}_1^+\bar{d}_1+\epsilon_2\bar{d}_2^+\bar{d}_2
+\epsilon_3\bar{d}_3^+\bar{d}_3,
\label{htqdd}
\eeqa
where $\epsilon_1=\left[E-\frac{\Delta V}{2}-2|t|
cos\left(\frac{2 \pi \phi}{3}\right)
\right]$, 
$\epsilon_2=\left[E-\frac{\Delta V}{2}-2|t|cos\left(\frac{2 
\pi (\phi+1)}{3}\right)\right]$ and
$\epsilon_3=\left[E-\frac{\Delta V}{2}-2|t|cos\left(\frac{2 \pi (\phi-1)}{3}\right)
\right]$.
Notice that the eigenvalues of the triple dot Hamiltonian depend
on the magnetic field. For the TQD-leads coupling Hamiltonian, $H_{LD}$, we obtain
\beqa
\bar{H}_{LD}=&&\bar{t}_{L1}c_0^+\bar{d}_1+\bar{t}_{L2}c_0^+\bar{d}_2+
\bar{t}_{L3}c_0^+\bar{d}_3\crcr
&&+\bar{t}_{R1}c_1^+\bar{d}_1+\bar{t}_{R2}c_1^+
\bar{d}_1+\bar{t}_{R3}c_1^+\bar{d}_3+hc.
\label{hintbar}
\eeqa
The new tunneling elements $\bar{t}_{Lj}$ and $\bar{t}_{Rj}$ are given by 
\beqa
\bar{t}_{L1}&=&\frac{1}{\sqrt{3}}\left(t_{L1}+t_{L2}\right)\cr
\bar{t}_{L2}&=&\frac{1}{\sqrt{3}}\left(t_{L1}+e^{-2\pi i/3}t_{L2}\right)
=\bar{t}_{L3}^{*}\cr
\bar{t}_{R1}&=&\frac{t_{R3}}{\sqrt{3}}\cr
\bar{t}_{R2}&=&\frac{t_{R3}}{\sqrt{3}}e^{2\pi i/3}=\bar{t}_{R3}^{*}.
\label{tbar}
\eeqa
In Eqs. (\ref{hintbar}) and (\ref{tbar}) we have omitted the magnetic flux dependence of the 
tunneling matrix elements between the dots and the leads in order 
to simplify the notation. In fact, this dependence does not appear when the left
lead is connected only to one dot, case that we shall analyze with more detail later.

Defining the amplitudes $\bar{D}_i=\langle {\bf r}|\bar{d}_i^+|0\rangle$, the 
Schr\"odinger equation reads as

\beqa
\label{systqd}\left\{
\begin{array}{lllll}
t_L C_{-1}+(\epsilon_{L}-\varepsilon) C_0+\bar{t}_{L1} \bar{D}_1+\bar{t}_{L2} \bar{D}_2
+\bar{t}_{L3} \bar{D}_3=0\cr
\bar{t}_{L1}^*C_0+(\epsilon_1-\eps)\bar{D}_1+\bar{t}_{R1}^*C_1=0\cr
\bar{t}_{L2}^*C_0+(\epsilon_2-\eps)\bar{D}_2+\bar{t}_{R2}^*C_1=0\cr
\bar{t}_{L3}^*C_0+(\epsilon_3-\eps)\bar{D}_3+\bar{t}_{R3}^*C_1=0\cr
\bar{t}_{R1} \bar{D}_1+\bar{t}_{R2} \bar{D}_2
+\bar{t}_{R3} \bar{D}_3+(\epsilon_{L}-\Delta V-\varepsilon) C_1+t_L C_2=0
\end{array}\right. .
\eeqa

Eq. (\ref{systqd}) allows us to express the amplitudes $C_{1},C_2$ as a function of
$C_{-1}$ and $C_0$. In so doing, one has to substitute the expressions for $\bar{D}_i$
in terms of the amplitudes in the leads
 and write the resulting relations as in Eq. (\ref{ttransfer}).
To simplify the expressions, in all the following discussions we will fix 
$\epsilon_{L}=0$  and the energy scale such as $t_L=-1$, which implies that
the energy band of the leads is from $-2$ to $2$. Then,
\beqa
{\cal T}_{11}&=&\left[\frac{\bar{t}_{L1}\bar{t}_{R1}^{*}}{\varepsilon-\epsilon_1}+
\frac{\bar{t}_{L2}\bar{t}_{R2}^{*}}{\varepsilon-\epsilon_2}+
\frac{\bar{t}_{L3}\bar{t}_{R3}^{*}}{\varepsilon-\epsilon_3}\right]^{-1}
\cr\cr
{\cal T}_{12}&=&-{\cal T}_{11}\left[-\varepsilon+
\frac{|\bar{t}_{L1}|^2}{\varepsilon-\epsilon_1}+
\frac{|\bar{t}_{L2}|^2}{\varepsilon-\epsilon_2}+
\frac{|\bar{t}_{L3}|^2}{\varepsilon-\epsilon_3}
\right]\cr\cr
{\cal T}_{21}&=&{\cal T}_{11}\left[-\varepsilon+
\frac{|\bar{t}_{R1}|^2}{\varepsilon-\epsilon_1}+
\frac{|\bar{t}_{R2}|^2}{\varepsilon-\epsilon_2}+
\frac{|\bar{t}_{R3}|^2}{\varepsilon-\epsilon_3}
\right]\cr\cr
{\cal T}_{22}&=&\frac{1}{{\cal T}_{11}^*}
%\frac{t_R^{(1)}t_L^{*(1)}}{\epsilon-\epsilon_1}+
%\frac{t_R^{(2)}t_L^{*(2)}}{\epsilon-\epsilon_2}+
%\frac{t_R^{(3)}t_L^{*(3)}}{\epsilon-\epsilon_3}
-\frac{{\cal T}_{12}{\cal T}_{21}}{{\cal T}_{11}}
\label{telements}
\eeqa
Equation (\ref{telements}) is the central result of the paper. Nevertheless,
the expressions for the transmission and reflection coefficients using Eqs.
(\ref{telements}) and (\ref{trleft}) are still too lengthy, so we shall analyze several
particular cases.

In the following subsections, we study two particular cases that can be
handled analytically. To get a clear understanding of the main 
features of the transmission we will
consider the simplest case
where $t_{L2}=0$ and $t_{L1}=t_{R3}=t_{LD}$.
We shall further simplify the problem assuming zero bias voltage.

\subsection{Transmission on-resonance with a single level\label{Fanores}}
Let us consider first the situation where the incident energy is very close
to one of the levels, e.g., level 1.
Furthermore, we will assume that the other two levels are far way, i.e.,
$|t_{LD}|\sim |\epsilon_1-\eps|<<|\epsilon_2-\eps|, |\epsilon_3-\eps|$.
Under these conditions, the Hamiltonian (\ref{htot}) reduces to the Fano-Anderson 
model\cite{mahan_book_1990} of a localized state in the continuum. 
In this case, the tranmission close to level 1 is given by a
Fano like resonance\cite{mahan_book_1990}
\beqa
|T(\eps)|^2\propto \frac{\left(q\Gamma/2+\eps-\epsilon_1\right)^2}
{(\eps-\epsilon_1)^2+\Gamma^2/4},
\label{fano}
\eeqa
where $q$ is the Fano parameter and $\Gamma$ the width of the resonance
defined in Ref. ~\onlinecite{fano_pr_1961}.
If the three levels are far apart, each level will lead to one of this 
Fano-resonances with their respective central energy, Fano parameter and widths.
Although we have used an implicit notation in which the magnetic flux dependence 
is not apparent, we should emphasize that the main variation of the transmission
with the magnetic flux in this single-resonance regime is governed by the sinusoidal
variation of the single particle energy levels $\epsilon_i$,  Eq. (\ref{htqdd}),
 with $\phi$. 
 
\begin{figure}
\begin{center}
\includegraphics[angle=-90,width=0.9\linewidth]{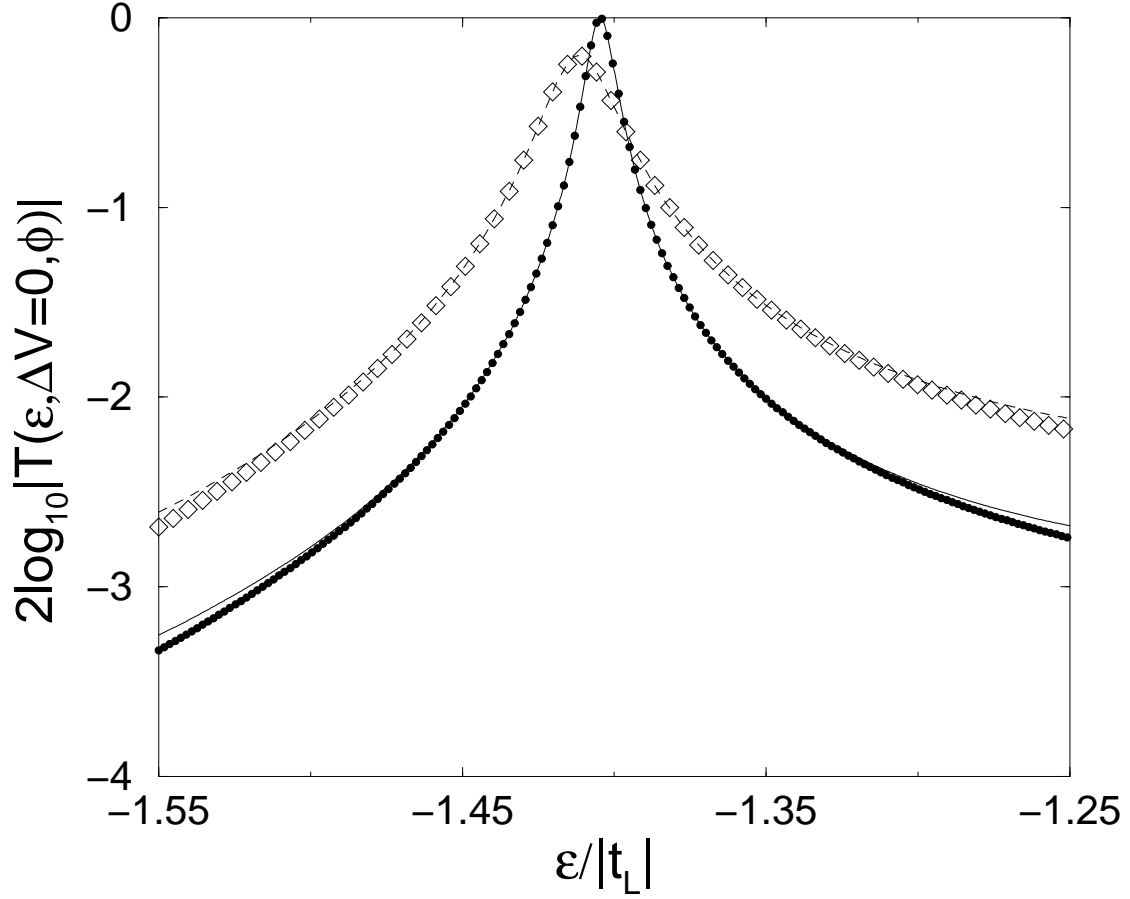}
\end{center}
\caption{$log_{10}$ of the transmission probability at zero magnetic field versus
the incident energy close to the on-resonance condition with the ground state
of the triple dot. Solid line corresponds to single lead-to-dot
connection and dashed line indicates the double connection.
Dots (diamonds) shows a fitting to a Fano line shape with
$q=86.73$ and $\Gamma=9.60\times 10^ {-3}$ 
($q=34.18$ and $\Gamma=2.40\times 10^ {-2}$). The other parameters are
$E=-1$, $\Delta V=0$, $t=-0.2$, $t_{L1}=t_{R3}=-0.05$ and $\phi'/\phi=1.73$
for the double connection.}
\label{tqdf_1_5}
\end{figure}

Fig. \ref{tqdf_1_5} shows the logarithm of the 
transmission probability versus the incident energy
at zero magnetic field for the single lead-to-dot connection (solid line) and the double
connection (dashed line), as well as the corresponding fitting to the form of Eq. (\ref{fano})
(dots and diamonds, respectively). For this case, where the coupling between the leads and the
dots is quite small compared to the tunneling $t_L$, the line shape is quasi-Lorentzian, as 
indicating the high values of $q$.

\subsection{Transmission close to a degenerate level\label{degenerate}}
\begin{figure}
%\begin{center}
%\includegraphics[angle=0,width=0.6\linewidth]{picture_sumary.eps}
\includegraphics[angle=0,width=0.6\linewidth]{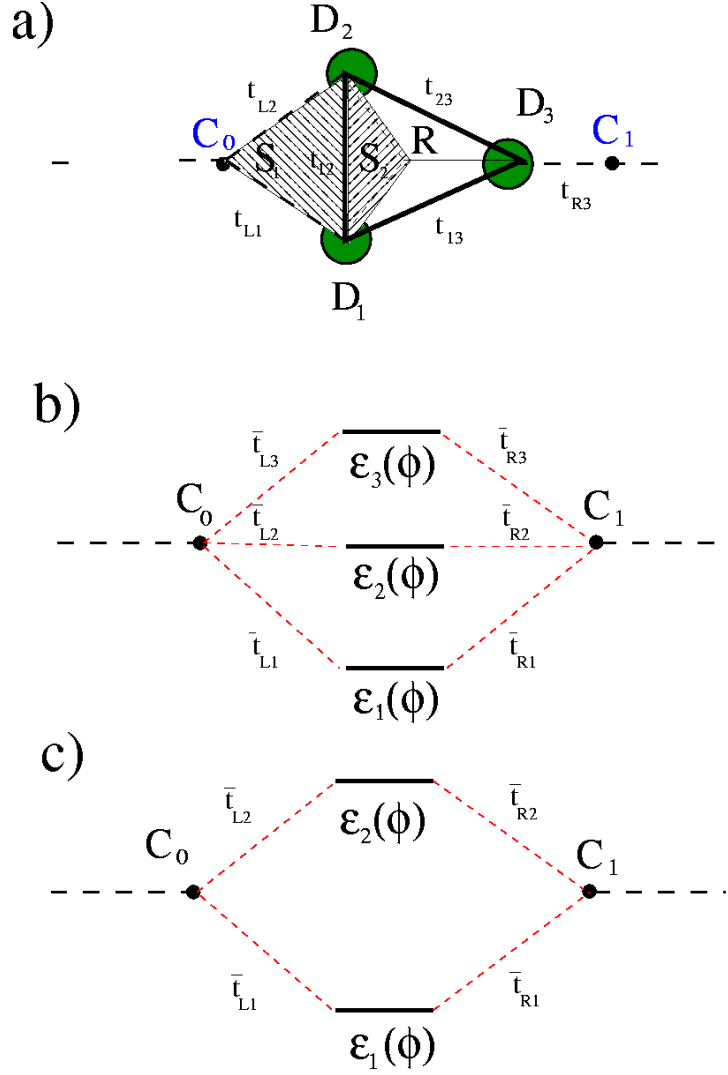}
%\end{center}
\caption{Schematic representation of the amplitudes and hopping matrix
 elements between sites: {\bf a)} 
in the original model, {\bf b)} in the bases of eigenvectors of the
isolated triple dot and {\bf c)}, the simplified version that accounts for
the case where the incident energy is close to a quasi-degenerate pair of levels
levels, i.e., $\varepsilon \approx \epsilon_1(\phi)\approx \epsilon_2(\phi)$.}
\label{tqdf_2}
\end{figure}

%\subsection{Dips in the conductance.\label{dips}}
%
Our aim now is to study the effects of the magnetic field
induced degeneracies 
of the triple dot on the transmission.
 When the energy of the incident 
electrons is close to the quasi degenerate level, the effect 
of the third orbital of the triple dot on the transmission can be neglected,
see lower panel of Fig. \ref{tqdf_2}. 
This approximation is valid for incident energies such 
$\tilde{E}-2|t_{LD}|\le \vep \le \tilde{E}+2|t_{LD}|$, where $\tilde{E}$
is the energy level of the degenerate states. The elements of the transfer
matrix can be obtained from Eq. (\ref{telements}) and, after the substitution
in Eq. (\ref{trleft}) and some extra algebra, the transmission probability
reads as
\beqa
&&\Big|T(\delta)\Big|^2=\cr\cr
&&\frac{(-1+e^{2ika})^2t_{LD}^4\delta^2}
{\beta\left[t_{LD}^2-3\delta(e^{ika}+\tilde{E}+\delta)\right]\left[t_{LD}^2-
\delta(e^{ika}+\tilde{E}+\delta)\right]\alpha(\delta) \left[\alpha(\delta)+2\delta+2e^{ika}
\delta(\tilde{E}+\delta)\right]},
\cr &&
\label{probte}
\eeqa
where $\beta =\left(-1+e^{i\pi/3}\right)\left(1+e^{2i\pi/3}\right)$ and 
$\alpha(\delta)=\delta+e^{ika}\left[-t_{LD}^2+\delta(\tilde{E}+\delta)\right]$. Here 
we have defined the energy shift $\delta=\vep-\tilde{E}$ and the corresponding
wavevector $k(\delta)a={\mbox arc \;cos}[(\delta+\tilde{E}) /2]$. Notice that we have written the 
previous expression in an apparently complex form, but it can be checked that Eq.
(\ref{probte}) provides a real positively defined quantity. Although this expression
is still quite complicated, it is clear that the transmission probability goes
to zero when we are on-resonance ($|T|^2\propto \delta^2$). This result was previously
described in the context of scattering through a tunneling junction with 
two resonant impurities in Ref.~\onlinecite{shahbazyan_raikh_prb_1994}.
 In fact, when the tunneling $t_{LD}$ is small enough,
i.e. $|t_{LD}|<<1$, and under the assumption $|t|<<1$, 
the transmission probability when the degenerate orbital level
is on-resonance with the Fermi energy of the leads ($\tilde{E}=0$),
 can be expressed as
\beqa
|T(\delta)|^2\approx \frac{2\delta^2 \Gamma(\delta)
 \frac{t_{LD}^4}{(t_{LD}^4+\delta^2)(t_{LD}^4+9\delta^2)}}
{\left(-\delta^2+\frac{\Gamma(\delta)^2}{4}\right)},
\label{tapprox}
\eeqa
where
\beqa
\Gamma(\delta) = \frac{2\left[t_{LD}^4-2(-5+2t_{LD}^2)
\delta^2+\frac{9\delta^4}{t_{LD}^4}\right]
t_{LD}^4}
{(t_{LD}^4+\delta^2)(t_{LD}^4+9\delta^2)}.
\nonumber
\eeqa
As we can see from Eq. (\ref{tapprox}), the transmission probability close to
the degenerate level $\tilde{E}$ can not be approximated by the addition
of two Fano resonances,
as one would naively expect from Eq. (\ref{fano}).

Let us analyze why the transmission coefficient goes to zero when the 
incident particles are on-resonance with a degenerate level. Let us
consider arbitrary (but small) tunneling elements 
$\bar{t}_{Lj},\;\bar{t}_{Rj}$ such that the two level approximation is still valid.
Without loss of generality, we assume that the degenerate levels are
$\epsilon_1$ and $\epsilon_2$. Then, the Schr\"odinger equation for the
incident energy $\vep=\tilde{E}=0$ can be written as
\beqa
\left\{ \begin{array}{lllll}
-C_{-1}+\bar{t}_{L1} \bar{D}_1+\bar{t}_{L2}\bar{D}_2=0\;\\
\bar{t}_{L1}^{*}C_0+\bar{t}_{R1}^{*}C_1=0 \\
\bar{t}_{L2}^{*}C_0+\bar{t}_{R2}^{*}C_1=0\\
-C_2+\bar{t}_{R1} \bar{D}_1+\bar{t}_{R2}\bar{D}_2=0\\
\end{array}\right. .
\label{onres}
\eeqa
The system of equations (\ref{onres}) admits two kinds of solutions depending on the
value of the determinant
\beqa
A=\mbox{det}\left(\begin{array}{cc}
\bar{t}_{L1}^{*} & \bar{t}_{R1}^{*} \\
\bar{t}_{L2}^{*} & \bar{t}_{R2}^{*}\\
\end{array}\right).
\label{deter}
\eeqa
Let us consider first the case where $A=0$. This implies that
$\bar{t}_{L1} \bar{t}_{R2}-\bar{t}_{R1}\bar{t}_{L2}=0$. Then,
making use of this relation in Eq. (\ref{onres}), one can extract the 
on-resonance transfer matrix
\beqa
{\cal T}(\vep=\tilde{E}=0)\equiv \left(\begin{array}{cc}
0 & -\bar{t}_{L1}^{*}/\bar{t}_{R1}^{*} \\
\bar{t}_{R2}/\bar{t}_{L2} & 0\\
\end{array}\right)
\eeqa
Using the relation between the transfer matrix ${\cal T}$ and the transmission, Eq.
(\ref{trleft}), one obtains
\beqa
T=\frac{2 i\bar{t}_{L1}^{*}\bar{t}_{R2}}
{\bar{t}_{L1}^{*}\bar{t}_{L2}+\bar{t}_{R2}\bar{t}_{R1}^{*}}.
\eeqa
If the tunneling elements differ only by a phase, 
$\bar{t}_{L1}=\bar{t}_{L2} \equiv \bar{t}e^{i\theta_L}$ and
$\bar{t}_{R1}=\bar{t}_{R2}\equiv \bar{t}e^{i\theta_R}$, with
$\theta_R,\theta_L, \bar{t}\in \Re$, then the only possible solution
is $|T|^2=1$ ({\em full transmission}).
This is in general the case of a double arm interferometer.

Now, we will consider the second case, $A\ne 0$. Notice that this is typically
the situation in Eq. (\ref{tbar}). The only possible solution of
the system of Eqs. (\ref{onres}) is then $C_0=C_1=0$, i.e.,
from the boundary conditions (\ref{boundary}) follows that $R=-1$ and $T=0$ 
({\em full reflection}).
If we assume a phase difference between the tunneling elements, i.e.,
$\bar{t}_{L1}=\bar{t}_{L2} \equiv \bar{t}$ and
$\bar{t}_{R1}=\bar{t}e^{i\theta_1};\quad \bar{t}_{R2}=\bar{t}e^{i\theta_2}$
with $\theta_i, \bar{t}\in \Re$, the amplitudes on the orbital levels 
1 and 2 are given by
\beqa
\bar{D}_1=\frac{-2ie^{i\theta_2}}{\bar{t}\left[1-e^{i(\theta_1-\theta_2)}\right]}\cr\cr
\bar{D}_2=\frac{2ie^{i(\theta_1-\theta_2)}}{\bar{t}\left[1-e^{i(\theta_1-\theta_2)}\right]}.
\label{amplitudes}
\eeqa
Notice that Eq. (\ref{amplitudes}) implies that the probability of finding the electron 
on each of the degenerate levels is the same

It is worth mentioning that the dips in the conductance are inherent to the 
two-channel resonant tunneling, \cite{shahbazyan_raikh_prb_1994} and they have been
described in double-dot Aharonov-Bohm 
interferometers
even at finite temperatures and in the presence of electron-electron 
interactions.\cite{kubala_konig_prb_2002,tokura_nakano_njphys_2007}

\section{Current and conductance\label{CG}}
To study the current through the system formed by the 
triple dot and the two leads we apply the Landauer-B\"uttiker 
formula.\cite{landauer_jresdev_1957,landauer_philosmag_1970,buttiker_prl_1986}
If the 
chemical potential of the left lead is $\mu_L$ and a bias voltage
$\Delta V/e=\left(\mu_L-\mu_R\right)/e$ is applied between the two leads, the 
current flowing through the system at zero temperature
is given by
\beqa
I(\Delta V,\phi)=\frac{e}{h} \int_{\mu_L-\Delta V}^{\mu_L}d\varepsilon\;|T(\varepsilon,\Delta V,\phi)|^2,
\label{current}
\eeqa
while the differential conductance can be obtained as
\beqa
G=\frac{\partial I(\Delta V,\phi)}{\partial \Delta V/e}=\frac{G_0}{2}\left[
\Big|T(\varepsilon=\mu_L-\Delta V, \Delta V,\phi)\Big|^2+
\int_{\mu_L-\Delta V}^{\mu_L}d\varepsilon\;\frac{\partial}{\partial \Delta V}|T(\varepsilon,\Delta V,\phi)|^2\right].
\label{conductance}
\eeqa
It relates the zero-temperature conductance to the transmission
probability $|T(\varepsilon,\Delta V,\phi)|^2$ at incident energy $\varepsilon$.
 Here $G_0=\frac{2e^2}{h}$
is the quantum of conductance. Notice that in the linear regime (small $\Delta V$),
the differential conductance (or just conductance) is
proportional to the transmission at the Fermi level
of the left lead, since the second term in Eq. (\ref{conductance}) cancels for 
$\Delta V\to 0$, while  at intermediate bias, the second term is responsible
of extra structure in the peaks
of the linear conductance, increasing the complexity of the 
profile as we increase the bias.\cite{castano_kirczenow_prb1990} 
The current and the conductance in the linear regime are
given by
\beqa
& I = \frac{e}{h}  |T(\varepsilon=\mu_L;0,\phi)|^2\Delta V,
\label{iapprox}\\
& G = \frac{G_0}{2} |T(\varepsilon=\mu_L;0,\phi)|^2.
\label{gapprox}
\eeqa
Therefore, the problem of obtaining the current
through the system is reduced to the calculation of the 
transmission coefficient
 $T(\varepsilon;\Delta V,\phi)$.

\section{Results\label{results}}
 Here we are interested in the regime in which
the energy band of the leads is much bigger than 
the energy splitting between the three levels of the TQD ($|t|<<1$).
Also, the tunneling 
between the dots and the leads will be taken much smaller than other
energy scales involved in the problem. 

\begin{figure}
%\begin{center}
%\includegraphics[angle=-90,width=0.9\linewidth]{tqd18_1.ps}
%\includegraphics[angle=-90,width=0.9\linewidth]{tqd18_2.ps}
\includegraphics[angle=-90,width=0.9\linewidth]{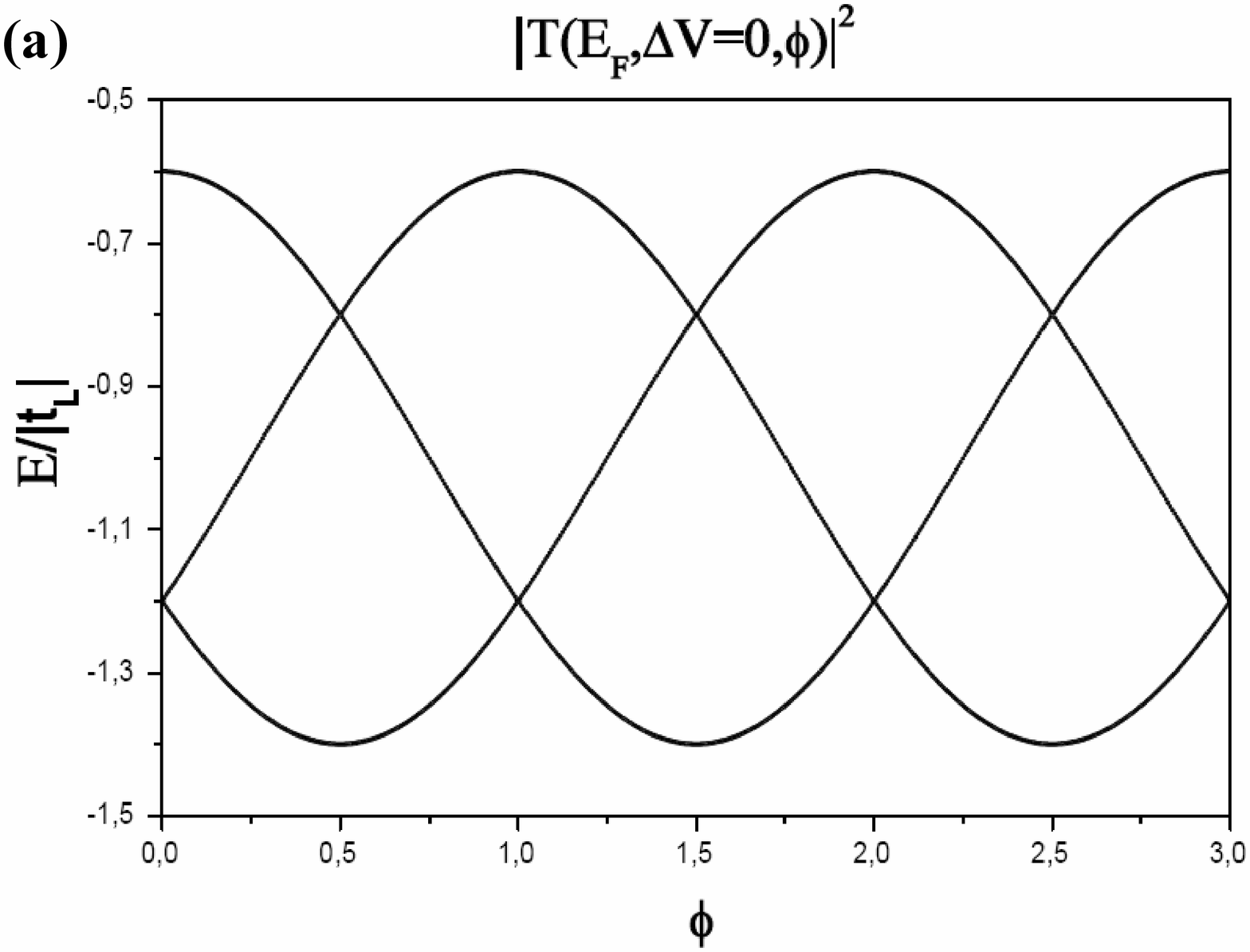}
\includegraphics[angle=-90,width=0.9\linewidth]{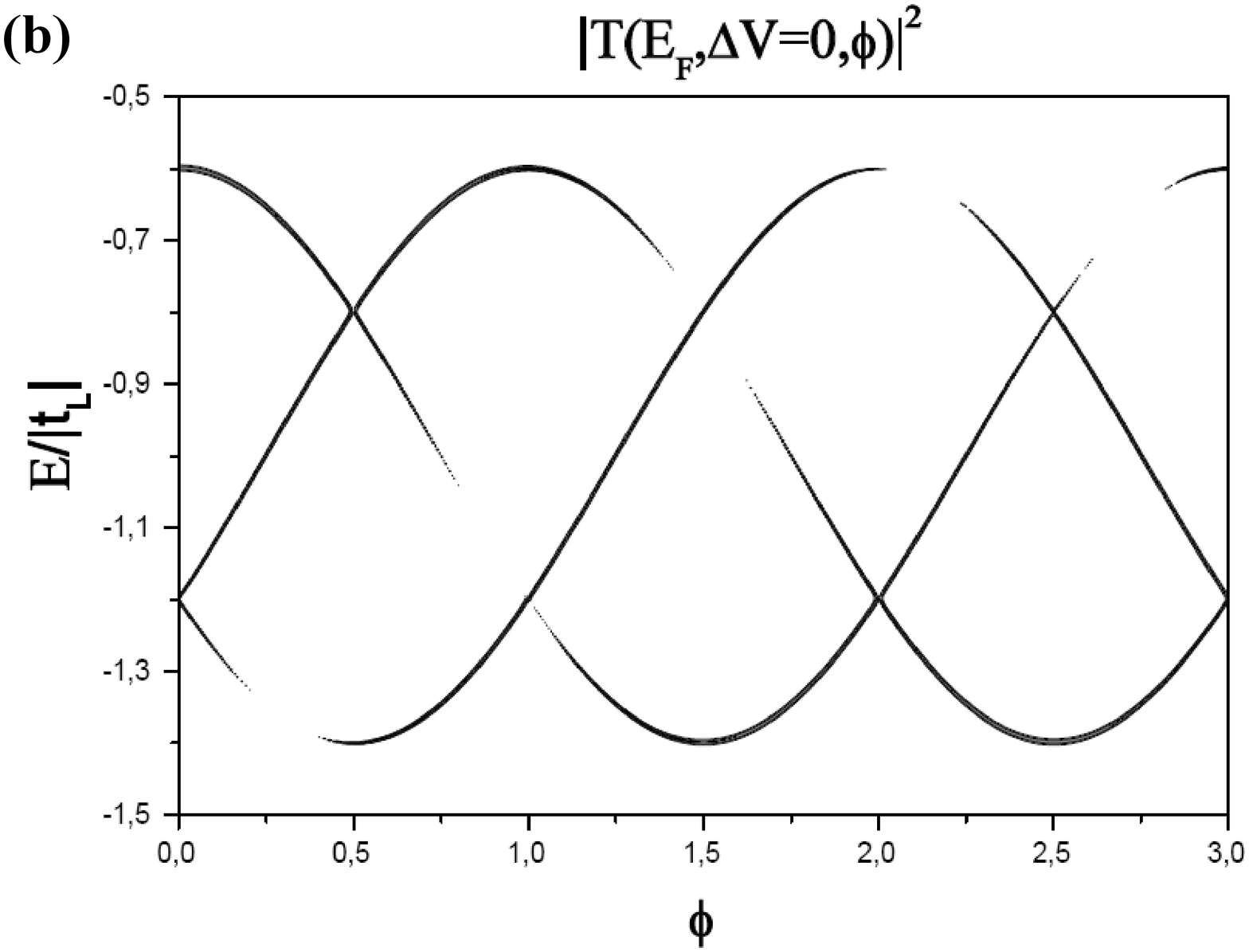}
%\end{center}
\caption{Transmission probability (vertical gray scale with black for 1) 
at the Fermi energy
versus the number of magnetic flux quanta $\phi$
and dot energy $E$ for the cases {\bf (a)}
 $t_{L2}=0$ showing periodic behaviour with $\phi$
and {\bf (b)}, $t_{L2}=t_{LD}$ with additional non-periodic structure.
 $t_{LD}=-0.05$, $\Delta V=0$ and $E_F=-1$.}
\label{tqdf_3}
\end{figure}

Let us consider first the linear transport, where 
Eqs. (\ref{gapprox}) and (\ref{iapprox}) are valid.
Fig. \ref{tqdf_3}(a) shows the transmission probability at the Fermi energy
 versus the magnetic flux and the dot energy $E$ for tunneling $t_{L2}=0$ with
$t=-0.2$, $t_{L1}=t_{R3}=t_{LD}=-0.05$, and $E_F=-1$.
 As shown in Fig. \ref{tqdf_3}(a), the transmission 
is periodic in the magnetic flux with period of one flux quantum. The transmission
pattern can be understood as follows. Electrons tunnel through
the TQD only when one of the three levels of the quantum molecule is on-resonance
with the Fermi energy ($E_F=-1$). For an arbitrary value of the magnetic flux,
this occurs for three different values of the dot energy $E$.
For example, at zero magnetic flux the resonance condition is fulfilled 
when the ground state
($E_G=E-2|t|$) is on-resonance
with the Fermi level ($E=-0.6$) while in the case of the doubly-degenerated
excited state, ($E_e=E+|t|$), this happens when $E=-1.2$. 
The oscillation in the levels of the isolated triple dot with the magnetic flux 
is reflected in the transmission since the values of high 
transmission correspond to dot energies 
on-resonance with the Fermi level. The structure that appears in Fig. \ref{tqdf_3}(a) 
is preserved for values of $|t_{LD}|\le |t|$, with the width of the
high transmission regions increasing with $t_{LD}$.
For $|t_{LD}|>|t|$,
the transmission is a
smoother function of the flux (not shown here) and the profile is 
deformed with respect to the case considered here. Fig. \ref{tqdf_3}(b) shows the 
transmission as a function of the magnetic flux and dot energy
 when the tunneling between the left lead and the second dot 
is allowed ($t_{L2}=t_{LD}$) for the ratio 
$\phi'/\phi =1.73 $. This ratio of fluxes leads to a non-periodic structure superimposed on the one
appearing in Fig. \ref{tqdf_3}(a), a consequence of the interference between
the two magnetic fluxes.

\begin{figure}
\begin{center}
\includegraphics[angle=-90,width=0.9\linewidth]{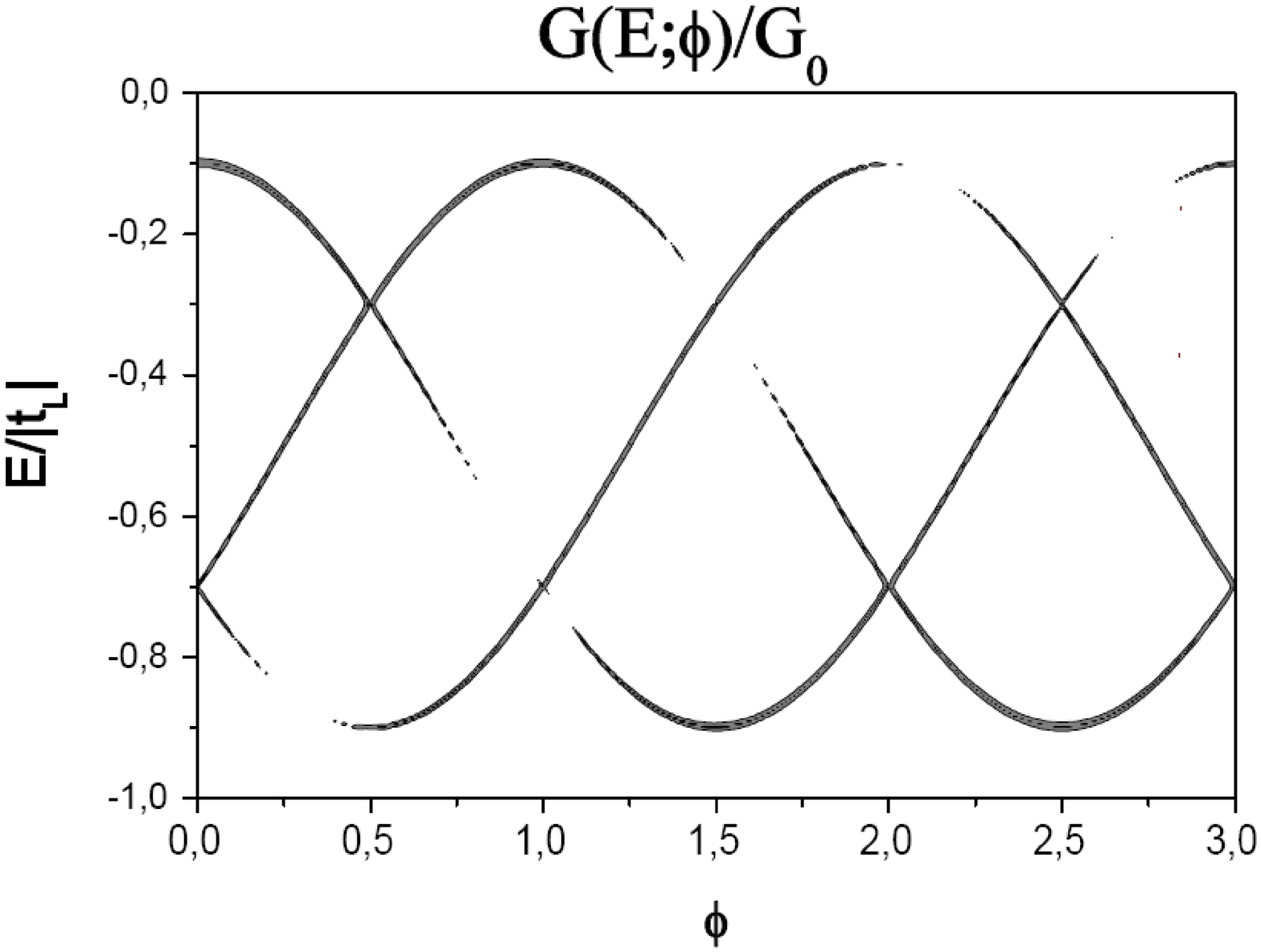}
\end{center}
\caption{Differential conductance $G$ (vertical gray scale)
versus number of magnetic flux quanta 
$\phi$ and the dot energy $E$ in the non-linear regime. 
$\Delta V=1$, $\mu_L=-1$, $t=-0.2$ and $t_{L2}=t_{LD}=-0.05$. The ratio of fluxes
is $\phi'/\phi=1.73$.}
\label{tqdf_4_0}
\end{figure}

Let us consider now a case in which the transmission
window given by $\Delta V$ is much bigger than $|t|$ (this is the 
case in most experimental setups with networks of lateral dots, including 
Ref.~\onlinecite{gaudreau_sachrajda_icps2006}). Then,
the total current contains contributions from all incident energies
within the transmission window, see Eqs. (\ref{current}) and(\ref{conductance}), and
corresponds to the non-linear regime.
 Fig. \ref{tqdf_4_0} shows a contour plot of the differential
conductance versus the magnetic flux and
the dot energy for $E_F=-1$ and $\Delta V=1$. 
The case depicted corresponds to $t_{L2}=t_{LD}=-0.05$. The variation of $G$ with the flux $\phi$
resembles the dependence of the transmission probability, shown in Fig. \ref{tqdf_3}(b),
allowing the determination of the tunneling matrix elements $|t|$ from the amplitude
of the oscillations. Therefore, the differential conductance 
under finite source-drain bias maps
 out the energy levels of the TQD.

\begin{figure}
%\begin{center}
%\includegraphics[angle=-90,width=3.in]{tqdp44ap.eps}
%\includegraphics[angle=-90,width=3.in]{tqdp44bp.eps}
%\includegraphics[angle=-90,width=3.in]{tqdp44cp.eps}
\includegraphics[angle=-90,width=3.in]{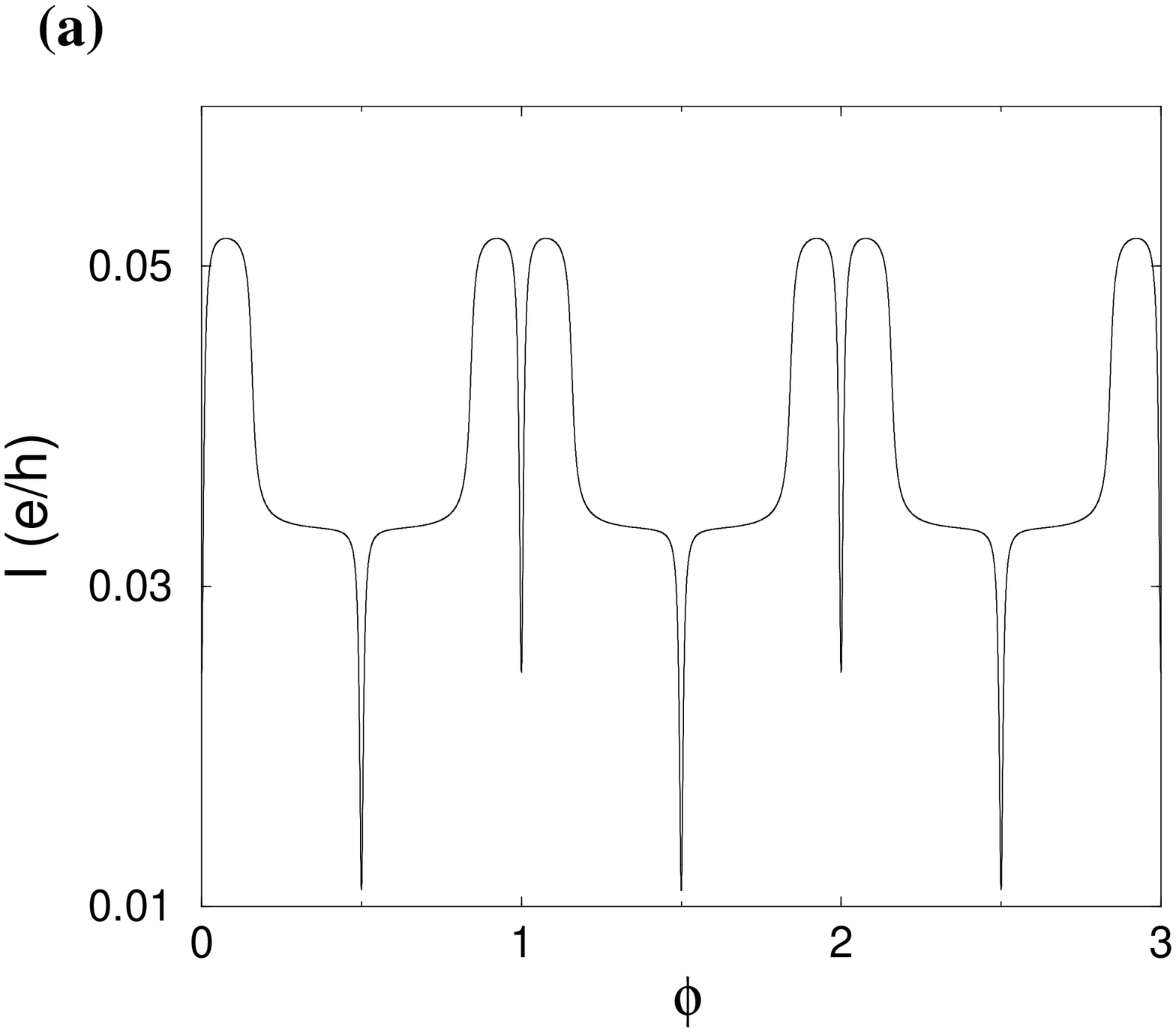}
\includegraphics[angle=-90,width=3.in]{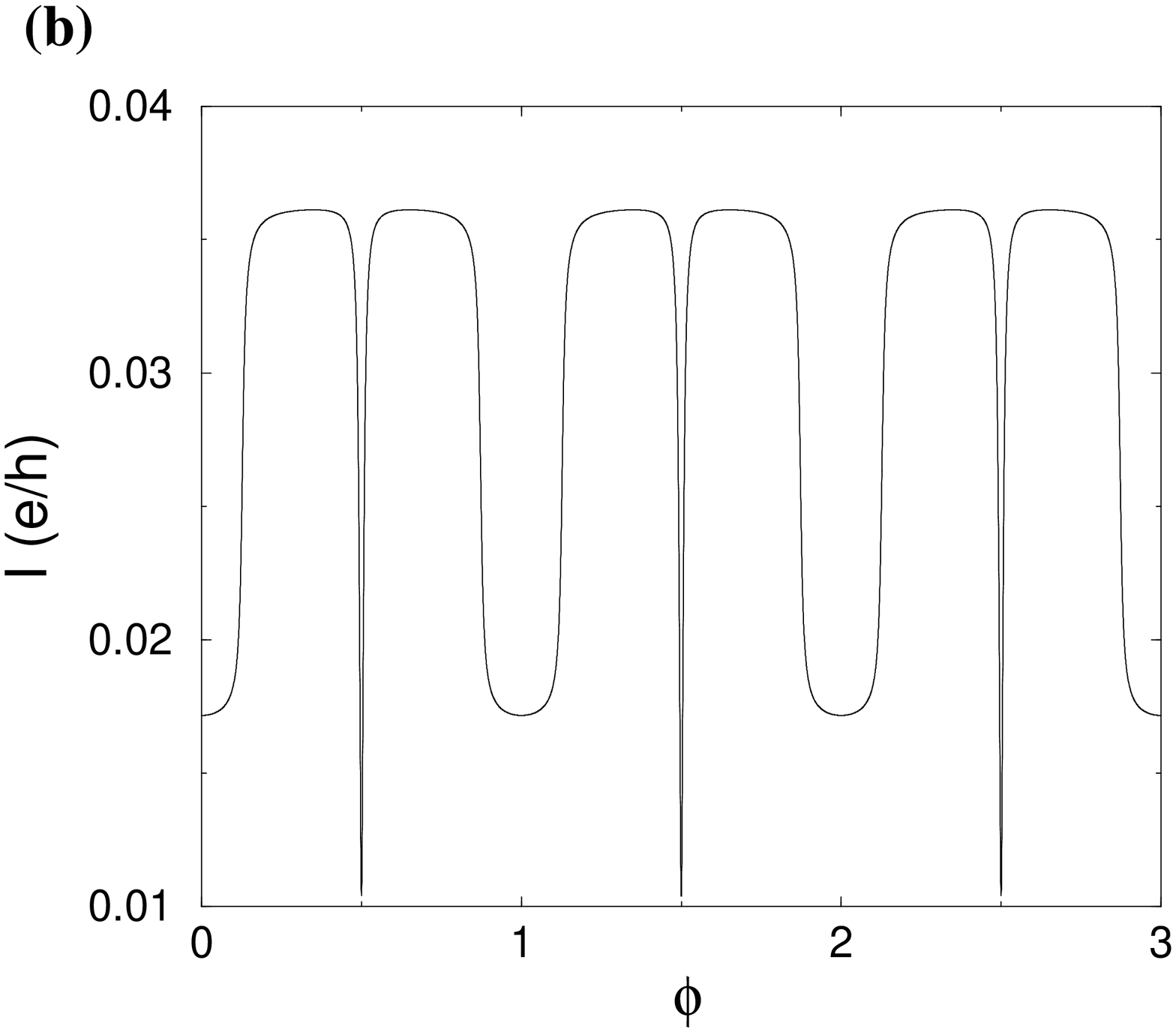}
\includegraphics[angle=-90,width=3.in]{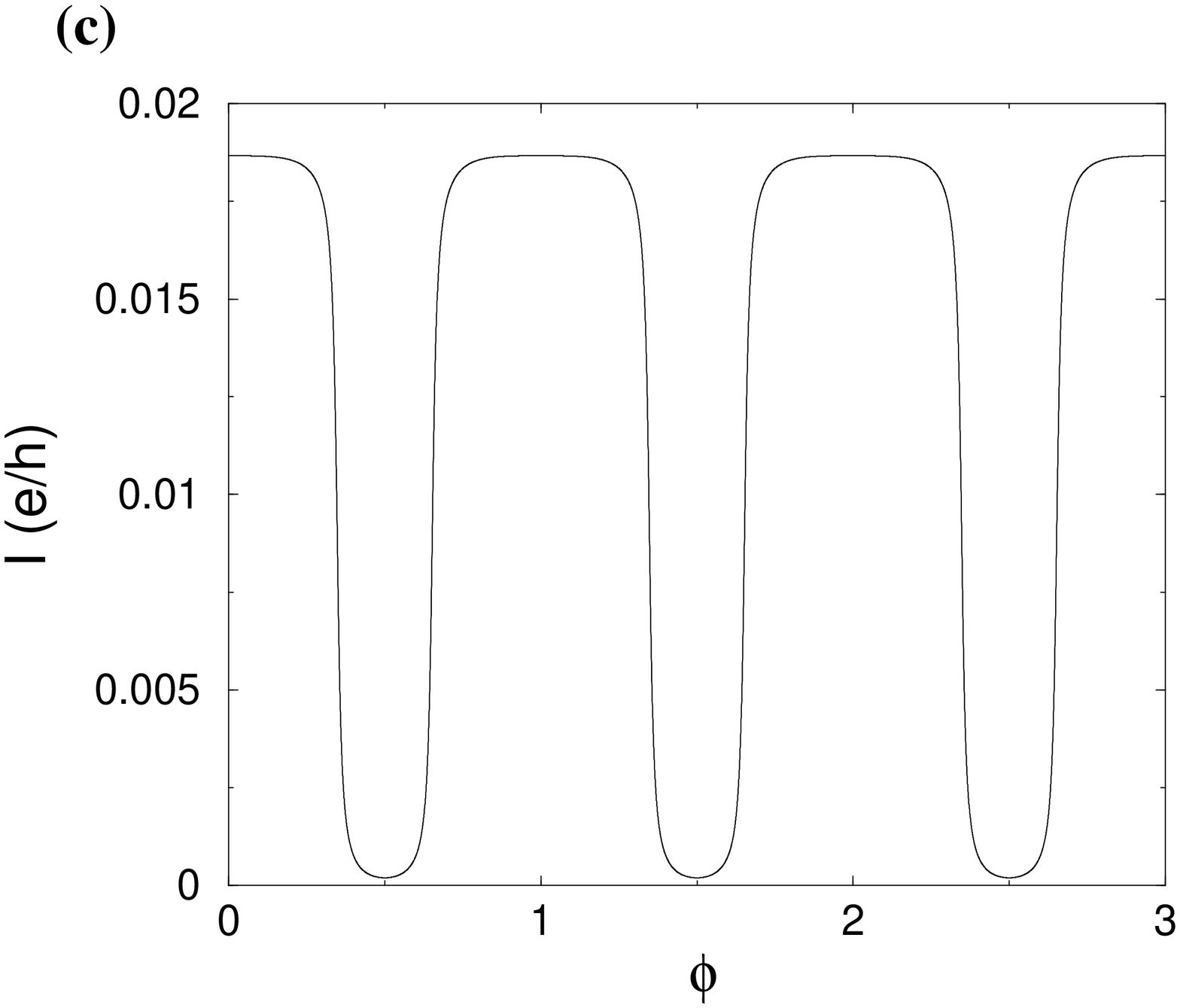}
%\end{center}
\caption{Current versus number of magnetic flux quanta $\phi$ in the non-linear regime 
under a bias $\Delta V=1$ and for the single lead-to-dot connection.
Fig. {\bf (a)}  shows the case where the three levels can contribute to the current, with
$E=-0.8$, Fig. {\bf (b)}, up to two levels ($E=-0.6$) and Fig. {\bf (c)} only one level
($E=-0.2$). $E_F=-1$, $t=-0.2$, and $t_{LD}=-0.05$. The periodicity of the drops in the current
changes for each case: $\phi_0/2$ in {\bf (a)}, $\phi_0$ in {\bf (b)} and no drops in {\bf (c)}.}
\label{fig_cluster}
\end{figure}

Although the contour plot of Fig. \ref{tqdf_4_0} provides the basic picture of the
behaviour of the TQD connected to the leads, 
it does not allow us to see several important details.
To simplify the analysis of the fine structure we can consider the simplest case
with single lead-to-dot connection and look at the current.
We have depicted the resulting current for three different values of the dot energy,
$E=-0.8$, Fig. \ref{fig_cluster}(a), $E=-0.6$, Fig. \ref{fig_cluster}(b), and
$E=-0.2$, Fig. \ref{fig_cluster}(c), using the same parameters as in Fig. \ref{tqdf_4_0}.
The first case, $E=-0.8$, corresponds to the scenario where the three levels of the triple dot
can contribute to the current. As we have shown previously, when the incident particle
has an energy on-resonance with degenerate levels, the transmission probability drops to zero.
In fact, even for the general case where $t_{L2}\ne 0$ and under an applied bias
$\Delta V$, it can be proven that close to the central energy $\tilde{E}$
\beqa
\left|T(\eps,\Delta V,\phi=\frac{n}{2})\right|^2\approx f(E,\Delta V)(\eps-\tilde{E})^2, \qquad \phi=\frac{n}{2},
\;n=1,2\dots
\label{taylorT}
\eeqa
where $f(E,\Delta V)$ is a function of the dot energy and the bias voltage. These zeros
in the transmission are reflected in the current as sharp drops whenever a pair of 
degenerate levels lies
between the chemical potential of the two leads. In Fig. \ref{fig_cluster}(a), this happens
when $\phi=n/2$, $n=1,2,\dots$. It should be pointed out that the zero
in the transmission probability at the degenerate level does not imply zero current,
since the large applied bias leads to a contribution from all three levels.
Anomalous behaviour in the transmission through a double-dot 
Aharonov-Bohm interferometer have also been 
described in Ref.~\onlinecite{kubala_konig_prb_2002,kubo_tokura_prb_2006}. In particular, 
Kubo {\it et al.} \cite{kubo_tokura_prb_2006} have reported sharp zero
conductance dips in the linear regime.
A second scenario appears in Fig. \ref{fig_cluster}(b), where
only two levels can contribute to the current.
 In this case, the anomalous dips in the current appear with a
periodicity of one flux quantum. Finally, the third possibility, at most one level contributing 
to the current, is presented in Fig. \ref{fig_cluster}(c). Here, the dips in the current have
disappeared since the possible degenerate states of the triple dot are outside the transmission window.

From our discussion, it should be clear that the anomalous behaviour of the
current with the magnetic field
is a manifestation of degeneracies in the system or, in other words, that
the presence of strong dependencies of the conductance with the magnetic field
is indicative of degeneracies in the system.

\section{Conclusions\label{conclusions}}
Summarizing, we have analyzed the linear and non-linear differential
conductance through an 
equilateral triple dot connected to two leads and subject to a perpendicular magnetic
field. Two possible spatial configurations were analyzed: a single lead-to-dot
connection where only one flux threads the system and a double connection
where two different fluxes must be considered. In both cases, we found
that superimposed on the AB oscillations induced by resonances with the oscillatory
levels of the TQD, sharp dips in the current appear whenever degenerate states 
lie between the chemical potential of the two leads. Therefore, three scenarios are possible:
no dips (degeneracies outside the transmission window), dips appearing
with a periodicity of one flux quanta (at most two level contributing to the current) and dips
with periodicity of half flux quantum (all three levels contributing).
We provided a simple theory of
the dips in the conductance. The presence of a double lead-to-dot connection produces
an additional non-periodic structure in the conductance 
as a function of the magnetic field, related
to the existence of two non-commensurate fluxes threading the system.
Both effects, AB oscillations and the dips in the current are also apparent
when large potential bias is applied between the two leads.

\section*{Acknowledgments}  
The Authors acknowledge support by the Canadian Institute for 
Advanced Research and QuantumWorks, and also useful discussions with  
A. Sachrajda, L. Gaudreau,  S. Studenikin, Y.-P. Shim and M. M. Korkusisnki.

\newpage

% ---------------- FIG. 0 BEGINS ----------------
% ---------------- END FIG. 0 ----------------  

% ---------------- FIG. 1 BEGINS ----------------
% ---------------- END FIG. 1 ----------------  

%---------------- FIG. 1_5 BEGINS ----------------
% ---------------- END FIG. 1_5 ----------------  

% ---------------- FIG. 2 BEGINS ----------------
% ---------------- END FIG. 2 ----------------  

% ---------------- FIG. 3 BEGINS ----------------
% ---------------- END FIG. 3 ----------------  

% ---------------- FIG. 4_0 BEGINS ----------------
% ---------------- END FIG. 4_0 ---------------- 

% ---------------- FIG. 4 BEGINS ----------------
% ---------------- END FIG. 4 ----------------  

\end{document}